\documentclass[twocolumn,POP, groupedaddress]{revtex4}
\usepackage{amsmath}

\usepackage{graphicx}          

\usepackage{times}

\usepackage{pifont}

\usepackage{color}

\def\simleq{\mathrel{\smash{\mathop{\raise2pt\hbox{$<$}}\limits_%
   {\smash{\raise4pt\hbox{$\sim$}}}}\vphantom\leq}}
\def\simgeq{\mathrel{\smash{\mathop{\raise2pt\hbox{$>$}}\limits_%
   {\smash{\raise4pt\hbox{$\sim$}}}}\vphantom\geq}}

\renewcommand{\vec}[1]{\ensuremath{\mathbf{#1}}}

\newcommand{\comment}[1]{}

\def\simleq{\mathrel{\smash{\mathop{\raise2pt\hbox{$<$}}\limits_%
   {\smash{\raise4pt\hbox{$\sim$}}}}\vphantom\leq}}
\def\simgeq{\mathrel{\smash{\mathop{\raise2pt\hbox{$>$}}\limits_%
   {\smash{\raise4pt\hbox{$\sim$}}}}\vphantom\geq}}

\renewcommand{\subsectionmark}[1]{\markright{\S~\thesubsection}}
\renewcommand{\sectionmark}[1]{\markright{\S~\thesection}}

\begin{document}

\title{Extension of the electron dissipation region in collisionless Hall MHD reconnection}
\author{Brian P. Sullivan}
\email{brian.sullivan@unh.edu}
\author{A. Bhattacharjee}
\email{amitava.bhattacharjee@unh.edu}
\author{Yi-Min Huang}
\email{yimin.huang@unh.edu}
\affiliation{Center for Integrated Computation and Analysis of Reconnection and Turbulence 
\\ Institute for the Study of Earth, Oceans, and Space\\
University of New Hampshire, Durham, NH 03824, USA}
\date{today}

\begin{abstract}
This paper presents Sweet-Parker type scaling arguments in the context of hyper-resistive Hall magnetohyrdodynamics  (MHD). Numerical experiments suggest that both cusp-like and modestly more extended geometries are realizable. However, the length of the electron dissipation region, which is taken as a parameter by several recent studies, is found to depend explicitly on the level of hyper-resistivity.  Furthermore, although hyper-resistivity can produce more extended electron dissipation regions, the length of the region remains smaller than one ion skin depth for the largest values of hyper-resistivity considered here--significantly shorter than current sheets seen in many recent kinetic studies. The length of the electron dissipation region is found to depend on electron inertia as well, scaling like $(m_e/m_i)^{3/8}$. However, the thickness of the region appears to scale similarly, so that the aspect ratio is at most very weakly dependent on $(m_e/m_i)$. 

\end{abstract}


\maketitle

\section{Introduction} 
The problem of fast magnetic reconnection in collisionless or weakly collisional plasmas has become a subject of great interest in recent years, in large part due to its relevance to impulsive phenomena such as magnetospheric substorms, solar and stellar flares, and the sawtooth crash in tokamaks. A key development has been the realization that non-ideal terms in the generalized Ohm's law (including electron inertia, electron pressure gradient, and Hall terms) provide a pathway to the onset of fast reconnection. This subject has a history spanning nearly two decades, with important contributions from both the fusion and space plasma communities (cf.~\cite{Bhattacharjee01, Bhattacharjee04} and references therein). It has been suggested that the Hall term in particular plays a key role in localizing the electron diffusion region. Within the framework of Hall MHD (or two-fluid) studies, there has been consensus on this point, though the issue of reconnection scaling  has remained controversial. Some studies of reconnection in Hall MHD systems have reported a ``universal'' normalized reconnection rate of $\sim0.1$, independent of dissipation mechanism \cite{Shay98b,Hesse99,Birn01,Shay01}, or system size\cite{Shay99,Shay04,Huba04}. However, other studies\cite{Grasso99,Wang01,  Porcelli02, Dorelli03a, Dorelli03b, Fitzpatrick04, Bhattacharjee05}  have found a broad range of dependencies on plasma parameters such as the ion skin and electron skin depths, and on boundary conditions, appearing to refute the claim of universality.

In an interesting sequence of studies based on fully kinetic simulations employing particle-in-cell (PIC) methods~\cite{Daughton06, Fujimoto06, Shay07}, questions have been raised regarding the conclusion, obtained from Hall MHD studies, that the electron diffusion region is highly localized.  In these PIC studies the electron diffusion region is not so localized. Rather it is elongated in the outflow direction to lengths on the order of $10 d_i$ ( where $d_i=c/\omega_{pi}$ is the ion skin depth) for electron to ion mass ratios of the order of $1/100$,  when Hall MHD simulations produce localized ($\le d_i$ length) electron current sheets (see e.g.~\cite{Rogers01, Ma01}). 
This raises the following interesting question: are fully kinetic simulations necessary to produce elongated electron current sheets? More specifically, is it possible to realize elongated electron current sheets within the framework of Hall or extended MHD models by means of a generalized Ohm's law that includes additional closure terms and parameterizes physical processes that may have kinetic origin? If the answer to this question is yes, it may facilitate greatly the capability of global multi-fluid models to represent the effects of kinetic physics at small scales.

A recent study by Chac\'on et al. \cite{Chacon07} presented scaling arguments and simulation results which indicated that in 2D electron MHD (EMHD) systems with electron viscosity (hyper-resistivity), the electron dissipation region can take on a wide range of geometries; the governing equations permit the aspect ratio (width/length) of the electron dissipation region to be of order unity (cusp like) or much smaller, i.e., leading to an extended electron current sheet. In the present work, we re-derive Chac\'on's EMHD scaling in perhaps a more transparent way, via Sweet-Parker type arguments about quantities at the upstream and downstream edges of the electron dissipation region. Similar approaches have been employed recently by Malyshkin~\cite{Malyshkin08}, and also by Uzdensky \cite{Uzdensky09}. These scaling results are then benchmarked numerically by two-fluid simulations of island coalescence both with and without electron inertia. We emphasize that neither this study, nor any other to date has produced a general method for analytically predicting the length of the dissipation region, which is perhaps the most important factor in determining whether the reconnection in a given system is fast or slow. We show, in particular, that the length of the electron diffusion region cannot be simply assumed to be a parameter fixed only by external boundary conditions, but depends on the mechanism that breaks the frozen-in condition (electron viscosity in this case). 

The organization of this paper is as follows. In section \ref{model}
we describe the simulation model and equilibrium profiles. In section \ref{scaling}, we derive our scaling arguments.  In section IV we describe the diagnostics used in analyzing simulation results. In section \ref{results} we present and discuss the results of the simulations included in this study. The main conclusions are summarized in section \ref{summary}.
\begin{figure}
\includegraphics[width=0.45\textwidth]{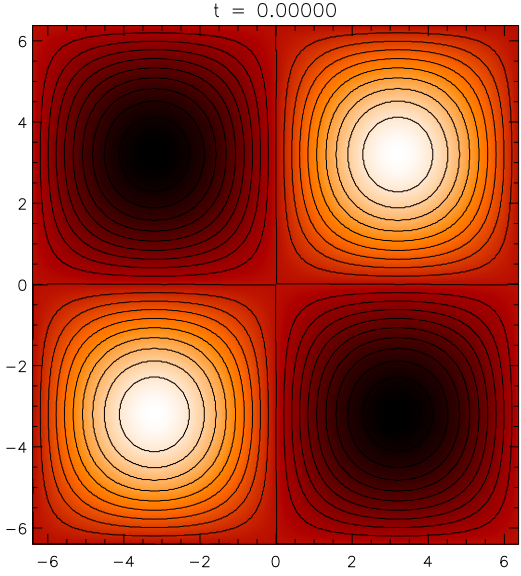}
\caption{(color online)
Initial out-of-plane current, $J_z(x,y)$ (orange scale), and magnetic field lines (black contours).
}
\label{initial_equilibrium}
\end{figure}
\section{Simulation Model}
\label{model}
\subsection{Model Equations}\label{Model Equations}
Our simulations are based on the two-fluid equations including  hyper-resistivity.
These equations in normalized form are \cite{Shay04}:
\begin{equation}
n\left(\partial_t\vec{V}_i+\vec{V}_i\cdot\nabla\vec{V}_i\right)
=\vec{J}\times\vec{B}-\nabla p,  
\label{mom}
\end{equation}
\begin{equation}
\partial_t\vec{B}'=-\nabla\times \vec{E}',
\label{faraday}
\end{equation}
\begin{equation}
\vec{E}' = - \vec{V}_i\times\vec{B}+\frac{1}{n}
\left(\vec{J}\times\vec{B}'-\nabla p_e \right) 
- \eta_H \nabla^2\vec{J}, 
\label{ohm}
\end{equation}
\begin{equation}
\partial_t n +\vec{V}_i\cdot\nabla n=-n\nabla\cdot \vec{V}_i,
\label{den}
\end{equation}
where $p_e=nT_e$ , $p_i=nT_i$ , $p=p_i+p_e$, 
$\vec{B}'=\left(1-d_e^2\nabla^2\right)\vec{B}$, and
$\vec{V}_e=\vec{V}_i-\vec{J}/n\ $, $\vec{J}=\nabla\times\vec{B}$.

For simplicity we assume an isothermal equation of
state for both electrons and ions (qualitatively similar to the
adiabatic case), and thus take $T_e$ and $T_i$ to be constant. 
The normalizations of Eqns.~(\ref{mom})-(\ref{den}) are based on constant
reference values of the density $n_0$ and the reconnecting component
of the magnetic field $B_{x0}$, and are given by 
(normalized $\to$ physical units): 
$t\to\omega_{ci}t$, 
$\omega_{ci}=eB_{0}/(m_ic)$,
$x\to x/d_i$, where
$d_{i,e}=c/\omega_{pi,e}$, and
$\omega_{pi,e}^2=4\pi n_0e^2/m_{i,e}$, 
$n\to n/n_0$, 
$\vec{B}\to\vec{B}/B_{0}$,
$\vec{V}_{i,e}\to \vec{V}_{i,e}/V_{A}$, where
$V_{A}=\omega_{ci}d_i=B_{0}/(4\pi n_0 m_i)^{1/2}$, 
$T_{i,e}\to T_{i,e}4\pi n_0/B_{0}^2$, 
$p_{i,e}\to p_{i,e}4\pi/B_{0}^2$, and
$\vec{J}\to\vec{J}/(n_0eV_{A})$.

Our algorithm employs fourth-order accurate spatial finite
differencing and the time stepping scheme is a second-order accurate
trapezoidal leapfrog \cite{zalesak1, zalesak2} . 
We consider a square, 2D slab with physical dimensions $L\times L = 12.8
d_i \times 12.8 d_i $ (so that $L=12.8$ in normalized
units) and periodic boundary conditions imposed at $x= \pm L/2$, and $y=
\pm L/2$.  
The simulation grid is $n_x \times n_y= 512 \times 512$, 
yielding grid scales of $\Delta_x=\Delta_y=0.025$. Note that compared to the simple form of Ohm's Law in resistive MHD, the generalized Ohm's Law in this two-fluid system (i.e. Eq.~(\ref{ohm}) above), contains four additional non-ideal terms: the Hall term ($\propto \vec{J}\times\vec{B}$), the electron pressure gradient term ($\nabla p_e$), the electron inertia terms, (which are manifested by the $d_e^2$ terms in the definition of $B'$), and the hyper-resistive term ($\propto \nabla^2\vec{J}$).  Note that of these non-ideal terms, only the hyper-resistive term is dissipative. Although hyper-resistivity (also known as electron viscosity) is generally negligibly small in collisional plasmas, it can be significant in a collisionless plasma. For a more thorough discussion of hyper-resistivity see, for example, Refs.~\cite{Schmidt71, Stix76, Strauss76, Kaw79, Strauss86, Boozer86, Hameiri87, Bhattacharjee95} . 
We consider the case of zero resistivity, and the model described here is thus intended to apply to collisionless plasmas.

\subsection{Initial Equilibrium}\label{Initial Equilibrium}
We begin with a two-dimensional (2D) equilibrium similar to that used by Longcope and Strauss (1994), shown in Fig~(\ref{initial_equilibrium}).  
The normalized magnetic field and density profiles in our initial equilibrium are given by:
\begin{align}
\vec{B}_x(x,y) &=  B_{0}
\left[ 
  \cos \left(\frac{2\pi y }{L}\right)\sin \left(\frac{2\pi x }{L}\right) 
\right] \vec{\hat{x}}, \\ 
\vec{B}_y(x,y) &=  -B_{0} 
\left[ 
\cos \left(\frac{2\pi x }{L}\right)\sin \left(\frac{2\pi y }{L}\right) 
\right]\vec{\hat{y}} , \\ 
n(x,y) &=  1+ \frac{1-B^2}{2(T_i+T_e)}.
\end{align}
The boundary conditions are periodic in both directions. As
required by these boundary conditions, $B$ is periodic under $y \to
y+L$ and $x \to x+L$ .
Note that the normalization parameter $B_{0}$ is equal to $2\pi/L=2\pi/12.8=0.4908$. The density profile is chosen to satisfy the total pressure
balance condition, which in normalized form is given by
\begin{equation}
n(T_i+T_e)+\frac{1}{2}B^2 ={\rm constant}.
\end{equation}
Unless otherwise stated, we take the (constant) total temperature to
be $T_{tot}=T_i+T_e=1.0$, or in physical units $4\pi
n_0T_{tot}/B_{x0}^2=1$, so that the plasma $\beta$ has a (minimum) value of 11.45 where $B_x=0.4908$ and
$n=1.3795$. The density reaches a maximum value of $n=1.5$ in the center of
the simulation domain $(x=0, y=0)$, where $B=0$. Initially
the current is carried entirely by the ions. To prevent energy buildup at the grid scale, the simulations
include fourth order dissipation in the density and momentum equations
of the form $\mu_4\nabla^4$ where $\mu_4=5.1\cdot10^{-7}$.  To avoid
physically artificial effects that can arise from exact reflection
symmetries of the initial condition, a small amount of random noise is
added to the magnetic field and ion current at the levels
$|\tilde{B}_{max}|\approx 10^{-5}$, $|\tilde{J}_{i_{max}}|\approx
10^{-5}$.

This initial equilibrium configuration contains four magnetic islands. Diagonally adjacent islands 
contain (out-of-plane) currents of like sign and are therefore mutually attracted. The islands with currents of opposite sign are of course mutually repelled. At rest these attractions and repulsions balance so that the equilibrium is metastable: either the upper right and lower left islands can coalesce at the center of the simulation domain, or the upper left and lower right islands can do so. Due to the symmetry of this system, whichever pair coalesces at the center, the opposite pair of islands will be repelled from the center coalescing at one of the corners of the simulation box. The symmetry of this metastable state is broken by giving the system an initial incompressible in-plane velocity perturbation of the form:
\begin{equation}
\vec{V}_i(x,y)=V_0 \left[ \sin\left(\frac{2\pi y }{L}\right)\vec{\hat{x}}  + \sin\left(\frac{2\pi x}{L}\right)\vec{\hat{y}} \right].
\end{equation}
Unless otherwise stated, the magnitude of this initial velocity is $V_0=0.1$ in normalized units.

\section{Scaling Arguments}
\label{scaling}
In systems where the Hall term is important, the non-ideal region takes on a two-scale structure: an outer region where the ions are demagnetized but the electrons are still frozen in, and an inner region where the electrons are demagnetized as well. It should be emphasized that in the outer part of the non-ideal region, although the ions are demagnetized, there is no dissipation occurring there. Here we will refer to this outer, non-dissipative zone as the ``Hall region.''  The inner, dissipative region, where neither species is magnetized, and where (in our case) electron viscosity is the dominant term, will be referred to as the ``electron diffusion region.''  Let $\delta_e$, and $L_e$ denote the thickness and length of the electron diffusion region, respectively.

In the following scaling arguments we neglect the effects of electron inertia, which is valid when the smallest scale of interest is much larger than the electron inertial length, $d_e$.  We also suppose that the spatial scale of the electron dissipation region is small sufficiently small compared to the ion inertial length, $d_i$, so that ion motion may be neglected. In other words, we consider an EMHD model. In the regime where the above conditions are met, the dominant terms on the right hand side of the generalized Ohm's Law (Eq.~(\ref{ohm})) are the Hall (proportional to $\vec{J} \times \vec{B}$) term, and the hyper-resistive term (proportional to $\eta_H \nabla^2 \vec{J} )$.  Let us consider the how these two terms conspire to influence the geometry of the electron region. Note that the arguments made here are very similar to those made by Uzdensky in a recently submitted paper~\cite{Uzdensky09}. 

From this point on we will work in a co-ordinate system rotated 45 degrees clockwise ($\hat{x} \to (1/\sqrt{2})({\hat{x}-\hat{y}})$) with respect to the system coordinates described in Section~(\ref{Initial Equilibrium}), so that we call the inflow direction ``$y$'' and the outflow direction ``$x$''.  In the case under consideration, there are two forces acting on the electrons as they flow away from the x-point: the $\vec{J} \times \vec{B}$ force accelerates the electrons away from the x-point along  the $x$ and $z$ directions in a channel of width $\sim \delta_e$, while the hyper-resistive (electron viscosity) term resists the shear in the electron fluid caused by the localized outflow. The peak electron outflow in steady state occurs at the point where these two opposing forces balance. This balancing of terms at the point of peak outflow is found to be well satisfied in our simulation results.  Here we will take the half length $(L_e/2)$ of the layer to be the distance from the x-point/stagnation point to the location of peak electron outflow. Let $B_{ey}$, $V_{ex}$, and $J_{ez}$ denote the magnetic field, electron outflow velocity, and out-of-plane current density at the point $(x,y)=(L_e,0)$. Then we have
\begin{align}
E_x(L_e/2,0) = \frac{B_{ey} J_{ez}}{c n e}  &- \eta_H \nabla^2 J_{ex} = 0, \nonumber \\ 
\implies \frac{B_{ey} J_{ez}}{c n e}           &\simeq  \eta_H \frac{n e V_{ex}}{\delta_e^2},  \nonumber \\
				    \implies  V_{ex} &\simeq \frac{B_{ey}J_{ez}\delta_e^2}{cn^2e^2\eta_H}, \label{eq1}
\end{align}
where the approximation that $\nabla^2 \sim \delta_e^{-2}$ holds provided that $\delta_e \ll L_e$. Between the inner and outer region, the electrons move at the $E \times B$ drift speed. So, at the downstream edge of the electron region, $B_{ey}= c E_z/V_{ex}$. In steady state, $\partial_t \vec{B} = \nabla \times \vec{E} = 0$ implies that $E_z$ is spatially uniform. Thus, $E_z(L_e,0)=E_z(0,0)=\eta_H \nabla^2 J_z(0,0) \simeq \eta_H B_{ex}/\delta_e^3$, (where $B_{ex}$ is the magnitude of the reconnecting component of the magnetic field at $(x,y)=(0,\delta_e)$) yielding:
\begin{equation}
B_{ey} \simeq \frac{c \eta_H J_{z0}}{V_{ex}\delta_e^2}.
\end{equation}
Substituting this value for $B_{ey}$ into Eq.~(\ref{eq1}) gives~\cite{Uzdensky09,Wang00}:
\begin{equation}
|V_{ex}| = |V_{ez}| = \frac{J_z}{ne} \sim \frac{1}{n e } \frac{c B_{ex}}{4 \pi \delta_e} = \frac{d_i V_A}{\delta_e}= \frac{d_e V_{Ae}}{\delta_e},
\end{equation}
where $V_A= B_{ex}/\sqrt{4 \pi m_i n}$, $V_{Ae}= B_{ex}/\sqrt{4 \pi m_e n}$, $d_{i}=c/\omega_{pi}$, and $d_{e}=c/\omega_{pe}$. We must emphasize here the quantity $(d_i V_{A})=(d_e V_{Ae})=(cB_{ex}/4\pi n e)$ is independent of the mass of either species. The present scaling arguments are essentially EMHD arguments, and as such there is no finite $d_i$. In the absence of electron inertia, there is also no $d_e$ in an EMHD model. 
Uniformity of $E_z$ holds \emph{across} the electron current sheet as well, requiring 
\begin{equation}
|V_{ey}| B_{ex} = \eta_H \nabla^2 J_{z0} \simeq \eta_H \frac{ B_{ex}}{\delta_e^3}.   
\end{equation}
Finally, taking the plasma to be incompressible (rigorously valid in the EMHD limit), requires that $|V_{ey}|L_e \simeq |V_{ex}| \delta_e$, we can solve for $\delta_e, |V_{ex}|,  |V_{ey}|$, and $E_z$ in terms of $V_A, d_i, \eta_H$, and $L_e$: 
\begin{align}
 \delta_e &= \left(\frac{\eta_H L_e}{d_i V_A}\right)^{1/3}, \label{delta_scaling}\\ 
|V_{ey}|  &= \frac{(d_i V_A)}{L_e}, \\
|V_{ex}|  &= |V_{ez}| = \frac{(d_i V_A)}{\delta_e} = \frac{(d_i V_A)^{4/3}}{(\eta_H L_e)^{1/3}}, \\
c E_z     &= \frac{(d_i V_A) B_{ex}}{L_e} = \frac{1}{L_e}\left( \frac{cB_{ex}^2}{4 \pi n e}\right) \label{Ez_scaling}
\end{align}
Equation~(\ref{delta_scaling}) predicts exactly the same scaling as that found by Chac\'on et al. (compare to Eq. (8) of reference \cite{Chacon07}). However, as in that study, the length of the electron dissipation region remains as a free parameter. This scaling permits (as stated by Chac\'on) both cases in which $\delta_e \sim L_e \sim \eta_H^{1/2}$, or $\delta_e\sim (\eta_H/ V_A)^{1/3} \ll L_e$. However, this statement may lead one to believe that there exists a bifurcated solution, which permits only the two aforementioned scalings. In fact, neither the arguments presented here nor those made in Refs.~\cite{Chacon07} or \cite{Malyshkin08} places any limit on the scaling of  $L_e$, thus failing to answer the question of whether the reconnection in any given system will be fast or slow. So in fact, a continuum of scalings may exist.
The reconnection rate given in Eq.~(\ref{Ez_scaling}) is not explicitly dependent on $\eta_H$, but may implicitly depend on $\eta_H$, depending on how $L_e$ scales.  Several limitations of the above scaling arguments are worth noting. First, the theory has assumed that ion motion is negligible on the spatial and temporal scale of the reconnection process. Second, the effects of electron inertia have been neglected. Third, electron pressure gradients have been assumed to be negligible, largely for the sake of simplicity; on scales $\gtrsim d_i$, electron pressure gradients are generally expected to be of the same order as the $\vec{J} \times \vec{B}$ term, as can be seen from Eq.~(\ref{ohm}). The first two of these points limit the applicability of these arguments to scales $\ell$ that satisfy the condition $d_e \ll \ell \ll  d_i$. A further, more subtle point, which was mentioned above, is that the downstream force balance criterion may not hold in all cases. However, even if there is no downstream force balance, the same scalings can be derived by enforcing steady state of the 2D EMHD equation. In two dimensions, the magnetic field can be expressed without loss of generality as $\vec{B}= \hat{z} \times \nabla \psi + B_z(x,y)\hat{z}$. In this case, the in plane and out-of-plane components of the EMHD equation can be expressed (in normalized form) as:
\begin{equation}
\begin{cases}
\partial_t \psi = -B_\perp \cdot \nabla_\perp B_z - \eta_H \nabla_\perp^2 J_z = - E_z = const.,\\
\partial_t B_z = -B_\perp \cdot \nabla_\perp J_z + \eta_H (\nabla_\perp^2)^2 B_z = 0. 
\end{cases} 
\end{equation}
In steady state, again taking $\partial_x \approx L_e^{-1}, \partial_y \approx \delta_e^{-1}, \nabla_\perp^2 \approx \delta_e^{-2}$, these two equations become:
\begin{equation}
\begin{cases}
\frac{B_{ex} B_{ez}}{L_e} \sim \frac{B_{ey} B_{ez}}{\delta_e} \sim \eta_H \frac{B_{ex}}{\delta_e^3},\\
\frac{B_{ex}^2}{\delta_e L_e} \sim \eta_H\frac{B_{ez}}{\delta_e^4},
\end{cases}
\end{equation}
and eliminating $B_{ez}$ from these two equations, again yields: $\delta_e = (\eta_H L_e / B_x)^{1/3}$, the same scaling given by Eq.~(\ref{delta_scaling}), and equivalent scalings come about for the other quantities as well. This approach is perhaps more general, as it doesn't require force balance at the downstream edge, though it may be less physically transparent. A general theory which predicts the length of the dissipation region in collisionless Hall MHD is in many ways the ultimate pursuit of all theoretical work on reconnection today, and as of yet no theory has delivered that result, except in special cases~ (e. g. \cite{Wang01,Shay07}). Lacking such a general theory, we now to numerical experiments to explore the dependence of $L_e$ on $\eta_H$.
\begin{figure}[t]
\includegraphics[width=0.45\textwidth]{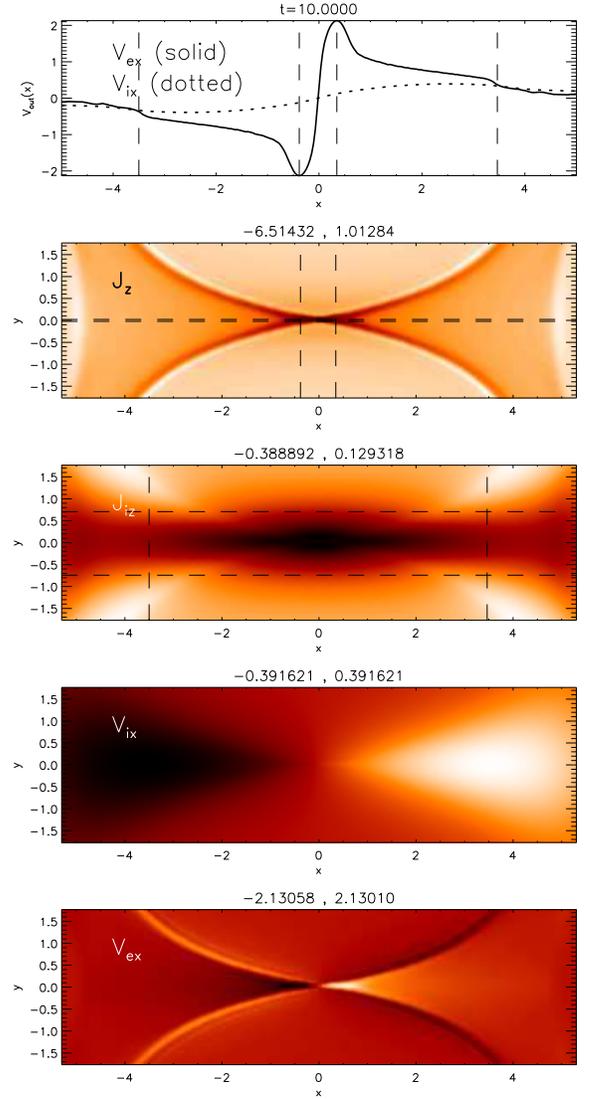}
\caption{
(color online) Dissipation region in a simulation with $m_e=0$. {\bf top panel:} 1D cuts of electron and ion outflow velocities, $V_{ex},\  \& \ V_{ix}$. The vertical lines mark the location of peak electron outflow (ends of electron diffusion region) and the points where electron and ion velocities come together. {\bf second panel:} $J_z(x,y)$; dashed lines mark edges of electron diffusion region. {\bf third panel:} $J_{iz}(x,y)$; dashed lines mark the locations where the ion and electron flows come together. {\bf fourth panel:} ion outflow,$V_{ix}(x,y)$, {\bf bottom panel: } electron outflow, $V_{ex}(x,y)$
}
\label{outflow_cut_t10.0000_r108}
\end{figure}
\section{Description of Diagnostics}
\begin{figure}[t]
\includegraphics[width=0.45\textwidth]{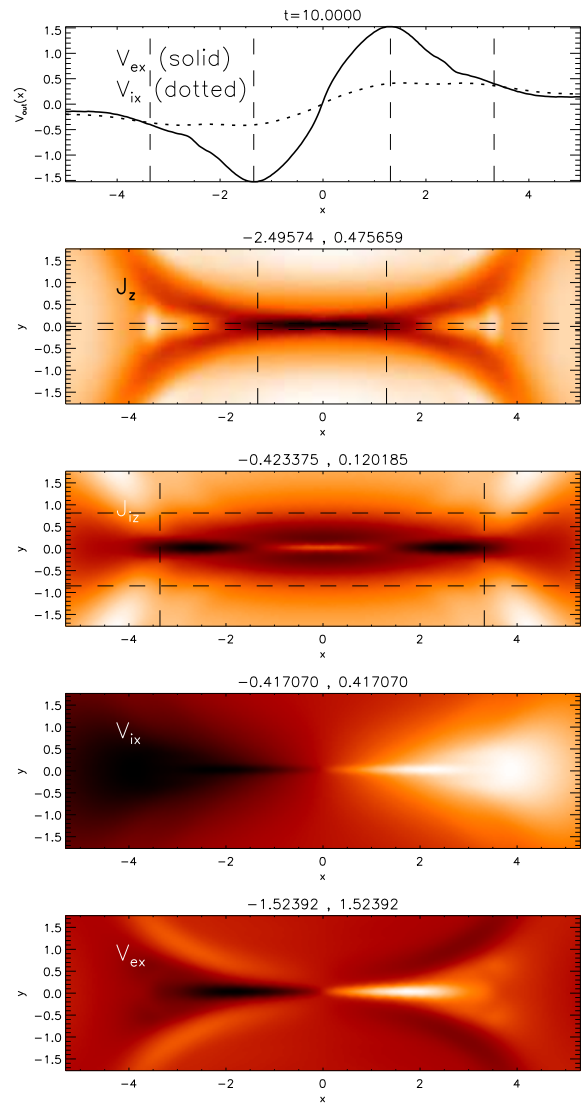}
\caption{(color online)
Dissipation region in a simulation with $m_e/m_i=1/25$. {\bf top panel:} 1D cuts of electron and ion outflow velocities, $V_{ex},\  \& \ V_{ix}$. The vertical lines mark the location of peak electron outflow (ends of electron diffusion region) and the points where electron and ion velocities come together. {\bf second panel:} $J_z(x,y)$; dashed lines mark edges of electron diffusion region. {\bf third panel:} $J_{iz}(x,y)$; dashed lines mark the locations where the ion and electron flows come together.  {\bf fourth panel:} ion outflow,$V_{ix}(x,y)$, {\bf bottom panel: } electron outflow, $V_{ex}(x,y)$
}
\label{outflow_cut_t10_r113}
\end{figure}

Before describing the results of our numerical experiments, we must describe the diagnostics employed in quantifying the extent of the non-ideal regions for each species.  Near the center of the current sheet, the out-of-plane current density, $J_z$, is due primarily to the electrons. In this study the width of the electron dissipation region is taken to be the full width at half max of the out-of-plane current, $J_z$. The downstream edge of the electron dissipation region is taken to be the location of the peak in the electron outflow speed. Outside of the non-ideal region, the two species flow roughly together at the $\vec{E} \times \vec{B}$ drift velocity. Both the upstream and downstream edges of the non-ideal dissipation region are taken to be the first point outside the electron dissipation region at which the electron and ion velocities differ by less than, say, ten percent.  For example, to determine the location of the downstream edge, we begin at the downstream edge of the electron dissipation region and search along $x$ until the quantity 
$\left(1- \frac{v_i(x,0)}{ v_e(x,0)}\right) \le 0.1$. Figure~(\ref{outflow_cut_t10.0000_r108}) shows the shape of the ion and electron dissipation regions at late time in a simulation with $m_e=0$. The top panel shows
1D cuts of $V_{ex}$ (solid), and $V_{ix}$ (dotted) along the outflow ($x$) direction. The inner pair of dashed vertical lines in this panel indicate the locations of peak electron outflow, which are taken to be the ends of the electron dissipation region. The outer pair of vertical lines mark the point where $\left(1- \frac{v_i(x,0)}{ v_e(x,0)}\right) = 0.1$, i.e. the ends of the Hall region. The second panel from the top shows the out of plane current, $J_z(x,y)$, along with dashed lines indicating the extent of the electron dissipation region. The third panel shows $J_{iz}$, the out-of-plane ion current, with the dashed lines now indicating the extent of the Hall region in the x-y plane. The lower two panels show $V_{ix}$, and $V_{ex}$ in the x-y plane. For comparison, corresponding results from a simulation with $m_e/m_i=1/25$ are shown in Fig.~(\ref{outflow_cut_t10_r113}). The point of peak electron outflow is further downstream than that observed in the simulation with no electron mass. This extension of the current sheet was seen in an earlier study~\cite{Ma99}, but was not well explored. Note that the electron outflow jets extend well beyond the boundaries of the electron diffusion region, and into the boundaries of the non-dissipative, Hall region. This feature is consistent with results PIC simulations. A recent paper by Hesse et al. \cite{Hesse08} provides a detailed discussion of the non-dissipative nature of these outflow jets in the context of a kinetic system. The shape and extent of the {\it Hall} region are basically unchanged by varying $\eta_H$ and $m_e/m_i$. So we will focus here on the electron dissipation region.
In examining the scaling geometry of the electron dissipation region, we will focus on quantities defined at the centers of the upstream and downstream edges. The upstream field, $B_{ex}$,  upstream density, $n_{e}$, and electron inflow speed $V_{ey}$, are measured at the point $(x,y)=(0,\delta_e)$, while the electron outflow speed is measured at the point $(x,y)=(L_e, 0)$, which, as mentioned above is taken to be the location of the peak electron outflow.
\section{Numerical Results}
\label{results}
\subsection{Hyperresistive Scaling}
\begin{figure}
\includegraphics[width=0.45 \textwidth]{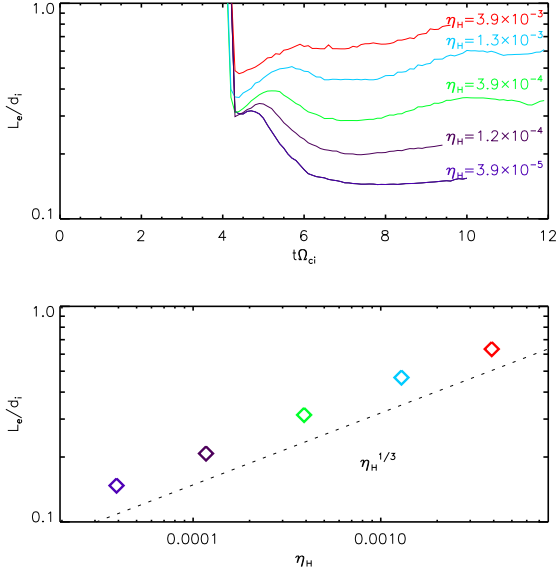}
\caption{(color online)
Time series of the length of the electron diffusion region, $L_e$, for varying levels of $\eta_H$ (upper panel), and scaling of median plateau values of $L_e$ vs. $\eta_H$ (lower panel). 
}
\label{hyper_resistive_length_scaling}
\end{figure}

\begin{figure}
\includegraphics[width=0.45\textwidth]{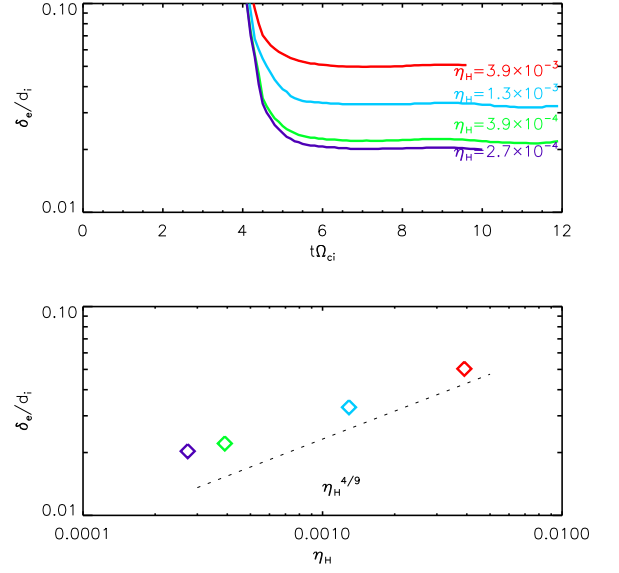}
\caption{(color online) Time series of the thickness of the electron diffusion region, $\delta_e$, for varying levels of $\eta_H$ (upper panel), and scaling of median plateau values of $\delta_e$ vs. $\eta_H$ (lower panel). 
}
\label{hyper_resistive_delta_scaling}
\end{figure}

\begin{figure}
\includegraphics[width=0.45\textwidth]{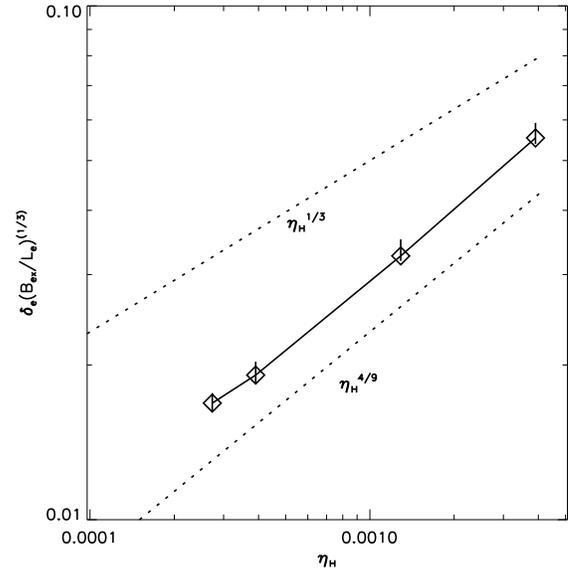}
\caption{
Scaling of $\delta_e(B_{ex}/L_e)^{1/3}$ vs. $\eta_H$}
\label{chacon_plot}
\end{figure}

\begin{figure}
\includegraphics[width=0.45\textwidth]{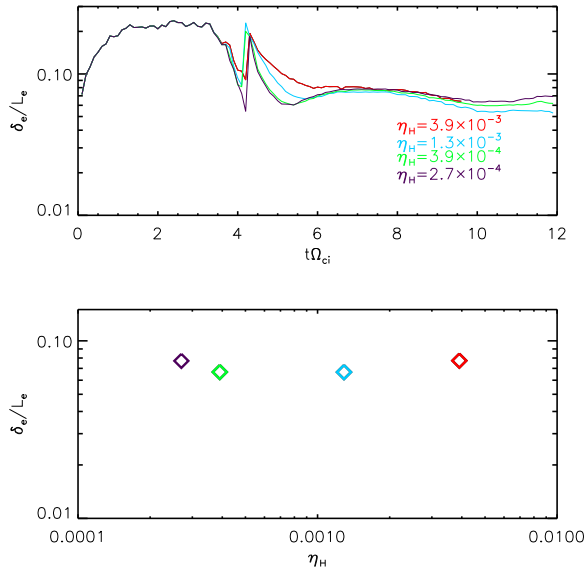}
\caption{(color online)Time series of the aspect ratio of the electron diffusion region, $\delta_e/L_e$, for varying levels of $\eta_H$ (upper panel), and scaling of median plateau values of $\delta_e/L_e$ vs. $\eta_H$ (lower panel). 
}
\label{hyp_aspect_ratio_scaling}
\end{figure}

\begin{figure}
\includegraphics[width=0.45\textwidth]{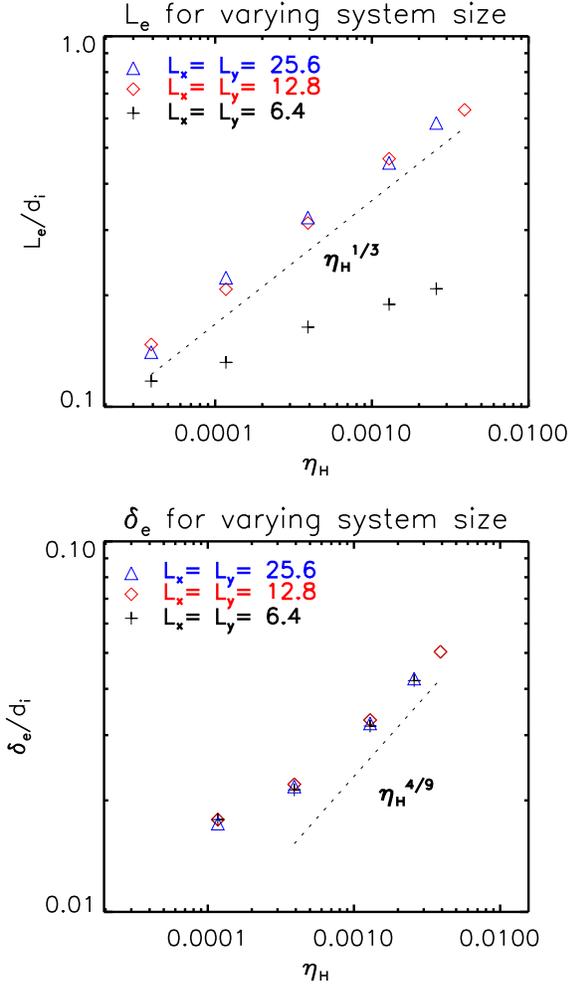}
\caption{(color online)Scaling of the Length of the electron diffusion region, $L_e$ vs. hyperresistivity, $\eta_H$ for varying system size(upper panel), and scaling of the electron diffusion region thickness, $\delta_e$ (lower panel). 
}
\label{multisize_hyp_scaling2}
\end{figure}

\begin{figure}
\includegraphics[width=0.45\textwidth]{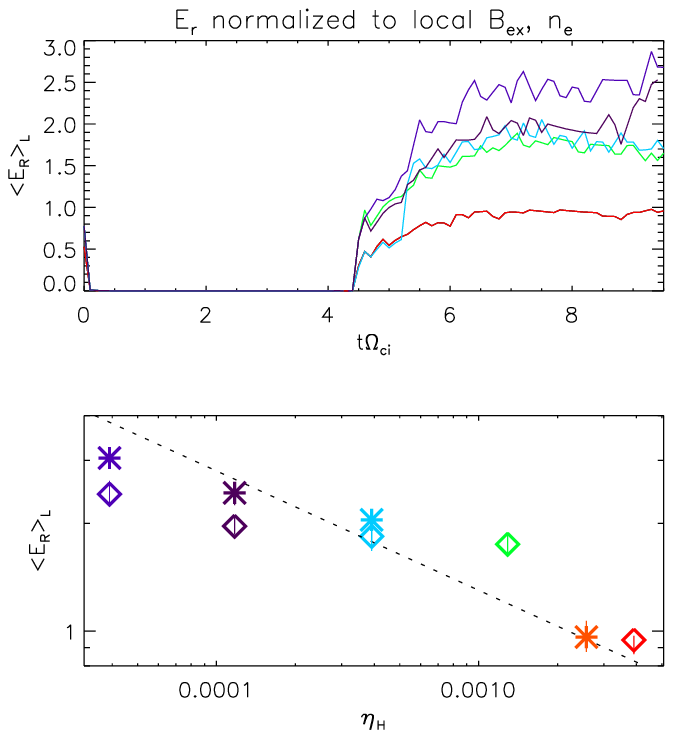}
\caption{(color online)Time series of the locally normalized reconnection rate, $\langle E_R \rangle_{L}$, for varying levels of $\eta_H$ (upper panel), and scaling of median plateau values of $\langle E_R \rangle_{L}$ vs. $\eta_H$ (lower panel)}
\label{rrate_hyp}
\end{figure}

\begin{figure}
\includegraphics[width=0.45\textwidth]{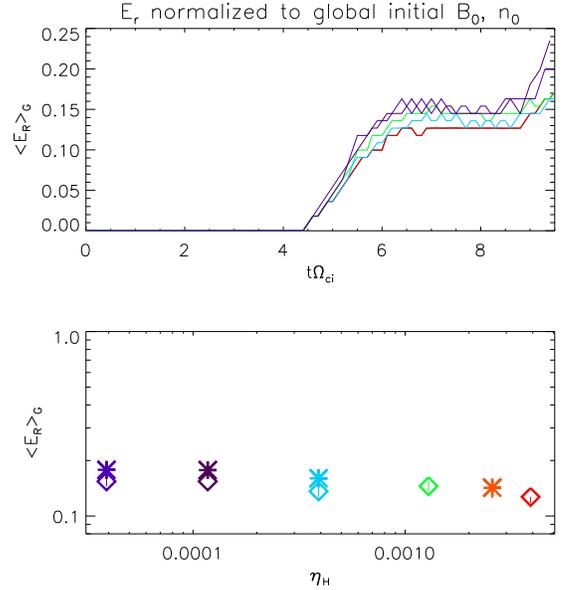}
\caption{(color online)Time series of the globally normalized reconnection rate, $\langle E_R \rangle_{G}$, for varying levels of $\eta_H$ (upper panel), and scaling of median plateau values of $\langle E_R \rangle_{G}$ vs. $\eta_H$ (lower panel)}
\label{rrate_hyp_global}
\end{figure}

Because, as mentioned above, no {\it ab initio} theoretical formulation of reconnection has yet successfully predicted the length of the current sheet in a Hall MHD system, we here begin our numerical investigation by describing the observed dependence of $L_e$ on varying hyper-resistivity, $\eta_H$. Figure~(\ref{hyper_resistive_length_scaling}) presents measurements  of the length of the electron dissipation region from five two-fluid simulations of magnetic island coalescence. Time series of $L_e$ are shown in the top panel of the figure. The five, solid curves represent the results of simulations with varying levels of hyper-resistivity, which do not include the effects of electron inertia. Notice that by $t \Omega_{ci} \sim 6.5$ each of these simulations has achieved a relatively stable value of $L_e$.  The lower panel of the figure plots the median value of $L_e$ from the plateau phase ($t \Omega_{ci} \ge 6.5$) versus $\eta_H$. 
The scaling of $L_e$ is quite clear, as shown by the dashed line (of slope $\eta_H^{1/3}$) in the lower panel. Empirically, $L_e \propto \eta_H^{1/3}$ in this modestly sized, periodic, coalescing system.
How does this empirical result for $L_e$ relate to the scalings predicted in the previous section? If one simply inserts $L_e \propto \eta_H^{1/3}$ into Eq.~(\ref{delta_scaling}), one finds $\delta_e \propto \eta_H^{4/9}$. Time series of $\delta_e$ are plotted in the upper panel of Fig.~(\ref{hyper_resistive_delta_scaling}). As was the case with $L_e$, the thickness of the electron dissipation region, reaches a stable value after $t \Omega_{ci} \sim 6.5$.  In the lower panel of Fig.~(\ref{hyper_resistive_delta_scaling}), the median plateau value of $\delta_e$ is plotted versus $\eta_H$. The dotted line, plotted for reference, represents a scaling of $\eta_H^{4/9}$, in reasonable agreement with the data points. In the case of the lowest value of $\eta_H$ presented in this figure, the current sheet thickness drops nearly to the grid scale. So that point may not be entirely reliable.
As a comparison to the results of  Chac\'on et al., in Fig.~(\ref{chacon_plot}) we plot $\delta_e(B_{ex}/L_e)^{1/3}$, analogous to Chac\'on's Fig. 3. The upper dotted line represents $\eta_H^{1/3}$, though the slope is actually closer to $\eta_H^{4/9}$ (as shown by the lower dotted line).
Note that if $L_e\sim \eta_H^{1/3}$ (purely an empirical result), and $\delta_e \sim \eta_H^{4/9}$ (in accordance with the predicted scaling of $\delta_e$, and the observed scaling of $L_e$), then the aspect ratio of the dissipation region has an \emph{extremely} weak dependence on hyper-resistivity $\delta_e/L_e \sim \eta_H^{1/9}$. Figure~(\ref{hyp_aspect_ratio_scaling}) presents the scaling of the  aspect ratio of the electron dissipation region. Time series of $\delta_e/L_e$ are plotted in the upper panel of the figure.  In the lower panel of the figure, diamonds represent the median plateau value of the aspect ratio. As expected , the aspect ratio varies extremely little across the range of hyper resistivities considered here. 
These scalings of $\delta_e$ and $L_e$ have been further tested in analogous systems of twice the physical size ($L_x = L_y =25.6$) and half the physical size ($L_x = L_y =6.4$) of the system described above. The results are shown in Fig.~(\ref{multisize_hyp_scaling2}). Not only the scaling of $L_e$, but even its absolute numerical value is practically unchanged by doubling the system size, as shown by the triangles and diamonds in the upper panel of Fig.~(\ref{multisize_hyp_scaling2}). The scaling of $L_e$ does however break down if the system size is too small, in which case the small size of the coalescing flux bundles limits the potential extent of $L_e$.  The dependence of $\delta_e$ on $\eta_H$, is unchanged by doubling or halving the system size. Thus for sufficiently large systems, the scalings found here do not appear to depend on the system size. This behavior appears to be quite different from that found in Ref.~\cite{Bhattacharjee05}, which used a different initial condition and did not achieve steady state at any point during the evolution of the system.
Next we turn our attention to the scaling of the reconnection rate. Time series of the electric field measurements are noisy. So, we consider measurements averaged over a window in time:
\begin{equation}
E_R \equiv \frac{\sqrt{n_e}}{B_{ex}^2} \left\langle \frac{\partial A_z}{\partial t}\right\rangle
\end{equation}
where $A_z$ is the component of the magnetic vector potential in the ignorable direction (i.e. the flux function up to a sign), and $\langle~\rangle$ denotes an average over an interval of one ion cyclotron time. Here the reconnection has been normalized by the the local upstream values of the magnetic field and density, $B_{ex}$, and $n_e$, which as mentioned earlier are measured at the point upstream where the out of plane current has fallen to half of its maximum value.  Figure~(\ref{rrate_hyp}) shows normalized reconnection rates computed from simulation results. Again the upper panel shows time series, while the lower panel shows median plateau values. These simulations  achieve a stable plateau value for $\langle E_R \rangle$ between $t\Omega_{ci}=6.5$ and $t\Omega_{ci}=8.5$. In the lower panel of Fig.~(\ref{rrate_hyp}) the diamonds represent the plateau value of normalized reconnection rate from the time series shown in the upper panel. The stars represent plateau values of the reconnection rate over the same range of hyper-resistivities in the larger ($L_x=L_y=25.6d_i$) system. The dotted line, plotted for reference represents a slope of $\eta_H^{-1/3}$, as predicted by Eq.~(\ref{Ez_scaling}) for our observed scaling of $L_e \propto \eta_H^{1/3}$. The observed reconnection rates appear to be within a factor of two of the predicted scaling.  We note that although it is common practice to normalize the reconnection rate using \emph{local} values of the magnetic field and density, these values are not necessarily values that fairly characterize the system. In this system for example, the meaningful physical purpose of the normalized reconnection rate is to characterize the time scale of the process of reconnecting a significant amount of the flux that was contained in the islands at $t=0$. To this end, it makes sense to use typical ``global'' values of magnetic field and density, which characterize the initial condition. Figure~(\ref{rrate_hyp_global}) shows reconnection rates from the same simulations as in Fig.~(\ref{rrate_hyp}). However, in Fig.~(\ref{rrate_hyp_global}) the reconnection rates have been  normalized to the global peak initial values of $B$ and $n$. Again the top panel shows time series, while the lower panel illustrates the scaling of the plateau value. These ``globally'' normalized reconnection rates appear to be nearly constant with a value of approximately $0.1-0.2$ across the range of hyper-resistivities considered here. Note that the ``raw,'' unnormalized reconnection rates
are also nearly independent of $\eta_H$, while the upstream field and density do appear to depend on the dissipation coefficient.  Figure ~(\ref{rrate_hyp_global}) makes clear that although the locally normalized reconnection rate depends on $\eta_H$, the time needed to reconnected a significant fraction of the initial flux in this system (as reflected by the ``globally'' normalized reconnection rate) is roughly independent of the dissipation region, and in this sense, the reconnection observed here is fast.

\subsection{Electron Mass Scaling}
We have observed above that the dimensions of the electron current sheet can be extended in cases which include electron inertia, as compared to hyper-resistive simulations that do not include the effects of finite electron mass (see Figs.~\ref{hyper_resistive_length_scaling} and \ref{hyper_resistive_delta_scaling}). Additionally, for a given mass ratio, there appears to be a threshold value of $\eta_H$ below which $L_e$ is determined by the mass ratio, and the effects of $\eta_H$ become negligible. The present study will not treat the general scaling of systems in which the effects of hyper-resistivity and electron inertia are comparable. However, we briefly present scaling results from simulations with varying electron mass, for $\eta=0, \eta_H=1.0\times 10^{-5}$. 
\begin{figure}
\includegraphics[width=0.45\textwidth]{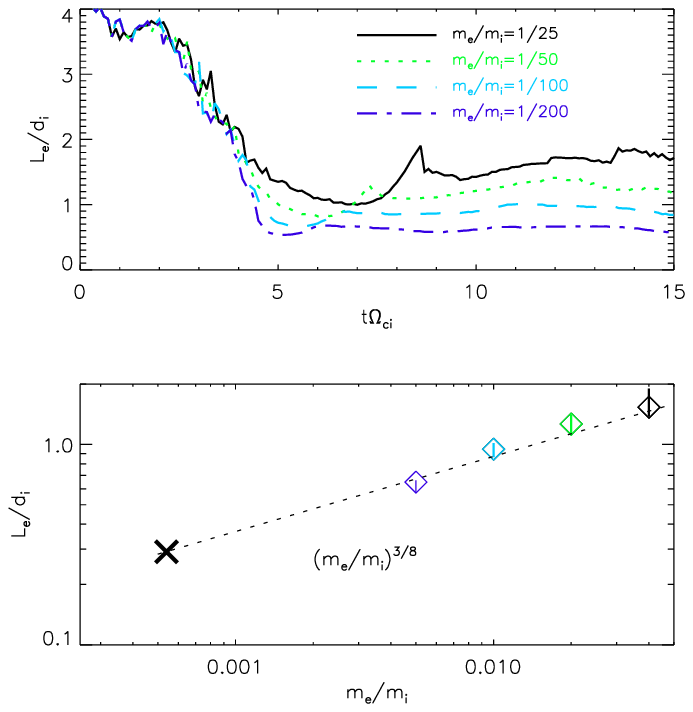}
\caption{(color online)Time series of the length of the electron diffusion region, $L_e$, for varying values of $m_e/m_i$ (upper panel), and scaling of median plateau values of $L_e$ vs. $m_e/m_i$ (lower panel). 
}
\label{e_mass_length_scaling}
\end{figure}
The top panel of Fig.~\ref{e_mass_length_scaling} shows time series of $L_e$ for mass ratios varying from $m_e/m_i=\tfrac{1}{25}$--$\tfrac{1}{200}$, and again the bottom panel plots the plateau values of $L_e$ vs. $m_e/m_i$. The dotted line in the lower panel represents an estimated scaling of $L_e \propto (m_e/m_i)^{3/8}$. This scaling is consistent with that found by Shay et al.~\cite{Shay07}, in the context of PIC simulations.  The X in the figure denotes an extrapolated value of $L_e\approx 15 d_e$ (or about 1/3 of an ion skin depth) for a physical electron/ion mass ratio of $1/1836$. Note that Shay et al. extrapolate to a value of $L_e\simeq0.6 d_i$ for a physical mass ratio, less than a factor of two different from our two-fluid result. This scaling can be understood using arguments presented in the aforementioned study by Shay et al.; the method is similar to that used earlier in this paper to derive hyper-resistive scalings. Suppose that the dominant terms in the electron momentum equation at the downstream edge of the electron dissipation region are the electron inertia term and the Hall ($\vec{J} \times \vec{B}$) term. Then in steady state ($\partial_t = 0$), and assuming the current to be entirely carried by the electrons at the scale of the dissipation region (the EMHD approximation) we have the following steady state electron momentum equation:
\begin{equation}
n m_e \vec{V}_e \cdot \nabla \vec{V}_e = n e (\vec{V}_e \times \vec{B}),
\label{steadystatemomentum}
\end{equation}
whose x component can be expressed as:
\begin{equation}
 \frac{\partial}{\partial x}\left(m_e\frac{V_{ex}^2}{2} \right) =  e V_{ez} B_y.
\end{equation}
This equation can be integrated from the x-point to the downstream edge to arrive at an equation for $L_e$. However, in order to integrate the above equation, we need to know something about the x dependence of $B_y$. In agreement with the PIC results of Shay et al.~\cite{Shay07} we have found that in our two-fluid model the profile of $B_y$ along the outflow direction in the dissipation region does not appear to depend noticeably on the mass ratio, as can be seen in Fig.~(\ref{By_compare}). Taking $V_{ez} \sim V_{Ae}$, assuming $B_y$ to vary linearly with distance from the x-point as $B_y(x)= B_y' x$, and integrating the above equation yields:
\begin{equation}
L_e \sim \left( \frac{m_e}{m_i} \right) ^{3/8} \left(\frac{E_z}{B_0 V_A} \right)^{1/2} \left( \frac{B_0}{d_i B_y'} \right)^{3/4}d_i
\end{equation}
This expression differs slightly from that obtained by Shay et al. They give the exponent of the $(B_0/d_i B_y')$ factor as $1$, whereas we find it to be $3/4$. Note that due to this difference our expression is entirely independent of $m_i$ (as it should be since there was no $m_i$ in Eq.~(\ref{steadystatemomentum}), whereas theirs contains an implicit $m_i$ dependence (via the $V_A$ in the second factor). This difference is not likely to become apparent in numerical experiments, as neither $B_0$ nor $B_y'$ appear to depend on the mass ratio. Biskamp et al. \cite{Biskamp97} have made different scaling arguments using the observed uniformity of the canonical momentum ($F=j_z-d_e^2 \psi$) along the upstream edge of the diffusion region, which yield a scaling of $L_e \sim (m_e/m_i)^{1/3}$. This scaling could also agree reasonably well with the results of our numerical experiment.  Resolving the difference between a scaling of $3/8$ and $1/3$ is difficult with only one decade of mass ratios. However, it is encouraging that our two-fluid scalings reasonably reproduce the mass ratio scaling seen in PIC simulations. 
\begin{figure}
\includegraphics[width=0.45\textwidth]{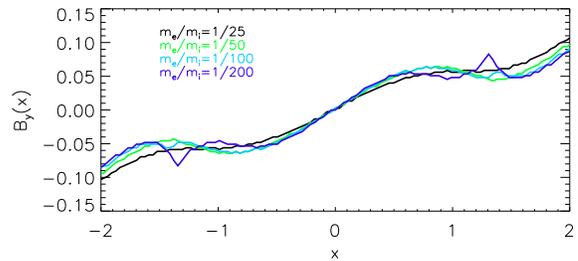}
\caption{(color online)Profiles of $B_y(x)$ along the outflow direction at  $t=11.0$ in simulations with varying mass ratios. Note that the slope of $B_y$ appears to be nearly independent of $m_e/m_i$ at x=0.
}
\label{By_compare}
\end{figure}

\begin{figure}
\includegraphics[width=0.45\textwidth]{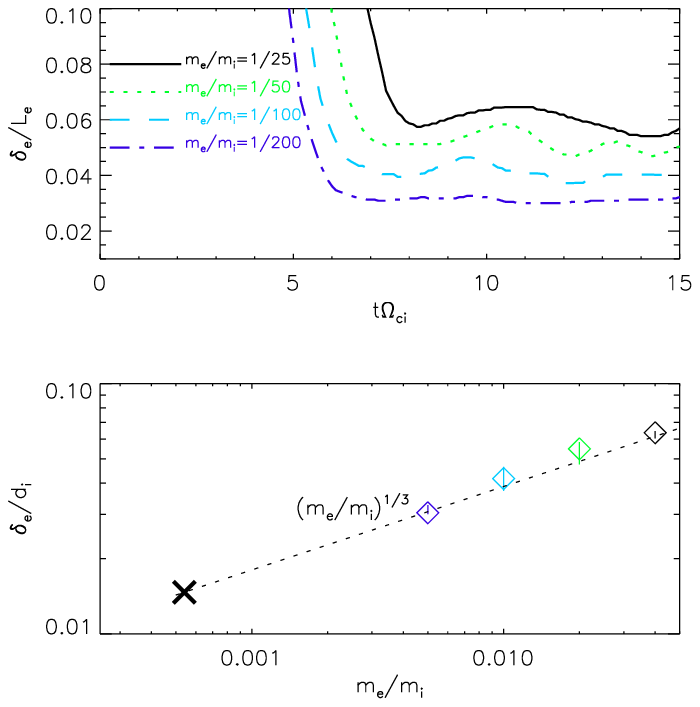}
\caption{(color online)Time series of the thickness of the electron diffusion region, $\delta_e$, for varying values of $m_e/m_i$ (upper panel), and scaling of median plateau values of $\delta_e$ vs. $m_e/m_i$ (lower panel). 
}
\label{e_mass_delta_scaling}
\end{figure}
The scaling of the dissipation region thickness, $\delta_e$, with $m_e/m_i$ is shown in Fig.~\ref{e_mass_delta_scaling}. The dotted line in the lower panel indicates a scaling of $(m_e/m_i)^{1/3}$. This is somewhat stronger than the scaling [$\delta_e/d_i \sim(m_e/m_i)^{1/4}$] observed by Daugton et al. \cite{Daughton06}, while showing good agreement with that observed by Dorfman et al. [$\delta_e/d_i \sim(m_e/m_i)^{1/3}$] in kinetic simulations of reconnection in MRX \cite{Dorfman08} (although in those simulations the frozen-in condition is broken by electron pressure agyrotropy rather than hyper-resistivity). Here the extrapolated valued of $\delta_e$ for a physical mass ratio is approximately $1 d_e$. The aspect ratio of the electron diffusion region does not appear to depend significantly on the mass ratio, since both the length and the width of the region are observed to scale with very similar powers of $(m_e/m_i)$.

\section{Summary}
\label{summary}
In this paper, we have made scaling arguments based on the balancing of terms in a hyper-resistive Ohm's law at the boundaries of the electron diffusion region, and have found a scaling for $\delta_e$ equivalent to that found by other recent studies\cite{Chacon07, Malyshkin08, Uzdensky09} via a somewhat different method. However, we emphasize here that these scaling arguments are ultimately incomplete unless they make a prediction about the length of the dissipation region. 
The expression derived for the reconnection rate (Eq.~\ref{Ez_scaling}) appears to be formally independent of electron mass, and hyperresistivity. However, formal independence is no guarantor of fast reconnection, because the upstream field and the length of the electron diffusion region may in fact depend on the dissipation mechanism. If $L_e$ is of order $d_i$ or even smaller then reconnection rate can take on a numerically high value, but numerical results indicate very clearly that $L_e$ is a function of $\eta_H$ for the simple coalescing system under consideration here. The reconnection rate normalized to local values of magnetic field and density depends on $\eta_H$ in this system, which might lead one to conclude that the reconnection is dissipation dependent. However, when the reconnection rate is normalized to typical \emph{global} values of $B$ and $n$, it appears that the time scale of the reconnection process is in fact nearly independent of $\eta_H$, and could therefore be termed ``fast''. For the purposes of comparison with other systems, these global values in general yield a more meaningful value for the normalized reconnection rate, since local values may themselves depend on the dissipation mechanism. 

Electron inertia does affect the extent of the electron diffusion region. However, both the length and width of the electron diffusion region appear to scale in very similar ways, so that the aspect ratio of the region appears to be independent of the mass ratio. The mass ratio dependence of the \emph{length} of the electron dissipation region found here is in good agreement with PIC results. The present study has made no attempt to test the scaling of reconnection with varying guide field, $B_z$, or boundary conditions, which are left for future work. Additionally, we have considered electron inertia and hyper-resistivity separately in our numerical experiments.  However, in the regime where both of these effects are significant, the parametric dependence of the reconnection region on both parameters could be complicated. 

Recent PIC simulations of reconnection have produced highly elongated current sheets, which break up into many small islands.  High Lundquist number simulations in the context of resistive MHD have also produced very extended current sheets, which are subject to a ``secondary tearing instability.'' This secondary instability also leads to the formation of many small plasmoids, similar to what is seen in the kinetic systems. The present study was largely motivated by the question of whether these kinds of extended electron current sheets could be realized within the framework of Hall MHD.  Electron viscosity does not on its own appear to be capable of producing highly extended current sheets. In collisionless kinetic systems, it has been shown that agyrotropic (off-diagonal) pressure tensor effects, which are not generally included in fluid models, play a dominant role in breaking the frozen-in condition. Hyper-resistivity models some of the physics associated with the electron pressure tensor. It is possible that a more complicated fluid closure parameterizing the physics of the agyrotropic pressure tensor components may be capable of reproducing the current sheet extension and subsequent plasmoid formation witnessed in kinetic systems. That investigation is a topic of presently ongoing research.\begin{acknowledgments}
This  work was supported by DOE Grant No. DE-FG02-07ER54832, NSF Grant No. ATM0802727, and NASA Grant No. NNX09AJ869.
\end{acknowledgments}


\end{document}